\documentclass[]{spie}  
\usepackage[dvips]{graphicx}

\title{Applying the Hilbert--Huang Decomposition to Horizontal Light Propagation $C_n^2$ data}

\author{Mark P. J. L. Chang, Erick A. Roura, Carlos O. Font\supit{1},
  Charmaine Gilbreath and Eun Oh\supit{2}
\skiplinehalf
\supit{1}Physics Department, University of Puerto Rico, Mayag\"uez, Puerto Rico 00680 \\
\supit{2}U.S. Naval Research Laboratory, Washington D.C. 20375
}

\authorinfo{Further author information: (Send correspondence to M.P.J.L.C.)\\M.P.J.L.C.: E-mail: mchang@uprm.edu, Telephone: 1 787 265 3844}

\begin{document} 
\maketitle 
\begin{abstract}
The Hilbert Huang Transform is a new technique for the analysis of
non--stationary signals.  It comprises two distinct parts: {\em{Empirical Mode 
Decomposition}} (EMD) and the {\em{Hilbert Transform}} of each of the modes
found from the first step to produce a Hilbert Spectrum.  The EMD is an
adaptive decomposition of the data, which results in the extraction of
Intrinsic Mode Functions (IMFs).  We discuss the application of the EMD
to the calibration of two optical scintillometers that have been used to
measure $C_n^2$ over horizontal paths on a building rooftop, and discuss the
advantage of using the Marginal Hilbert Spectrum over the traditional Fourier
Power Spectrum.
\end{abstract}


\section{INTRODUCTION}
\label{sect:intro}
The common practice when studying time series data is to invoke the tools of Fourier spectral analysis.  Although extremely versatile and simple, the technique suffers from some stiff contraints that limit its usefulness when attempting to examine the effects of optical turbulence in the frequency domain.  Namely, the system must be linear and the data must be strictly periodic or stationary.  Strict stationarity is a constraint that is impossible to satisfy simply on practical grounds, since no detector can cover all possible points in phase space. The linearity requirement is also not generally fulfilled, since turbulent processes are by definition non--linear.

Fortunately a new technique that has come to be known as the Hilbert Huang Transform (HHT) has been developed\cite{Huang1998}, patented by NASA.  This allows for the frequency space analysis of non--stationary, non--linear signals.  The HHT is composed of two main algorithms for filtering and analyzing such data series.  Firstly it employs an adaptive technique to decompose the signal into a number of Intrinsic Mode
Functions (IMFs) that have well prescribed instantaneous frequencies, defined as the first derivative of the phase of an analytic signal.  The second step is to convert these IMFs into an energy--time--frequency relationship, by means of the Hilbert Transform.  

Asides from overcoming the problems associated with more traditional Fourier methods, the HHT makes it possible to visualize the energy spread between available frequencies locally in time, rather like wavelet transform methods.  The advantage the HHT has over wavelet transforms is that it is of much higher resolution, since it does not {\em{a priori}} assume a basis; rather it "lets the data do the talking".

\section{INSTANTANEOUS FREQUENCY}
\label{sect:instfreq}
Key to the HHT is the idea of instantaneous frequency, which we will sometimes refer to as simply "the frequency".
The ideal instantaneous frequency is quite simply the frequency of the signal at a single time point.  No knowledge is required of the signal at other times.  Naturally such a statement leads to difficulties in definition; Huang et al\cite{Huang1999} take it to be the derivative of the phase of the analytic signal, found from the real and imaginary parts of the signal's Hilbert Transform, which we follow.

The immediate problem in dealing with a phase so defined is that, for the most part, the Hilbert Transforms of the direct signals are not well behaved resulting in negative instantaneous frequencies which do not represent physical effects.  The method by which this is circumvented is to ensure that the input to the Hilbert Transform obeys the following conditions:

\begin{itemize}
\item[(a)] The number of local extrema of the input and the number of its zero crossings must be either equal or differ at most by one.
\item[(b)] At any point in time $t$, the mean value of the upper envelope (determined by the local maxima) and the lower envelope (determined by the local minima) is zero.
\end{itemize}
The functions that obey these are considered the IMFs.

\section{EMPIRICAL MODE DECOMPOSITION} 
\label{sect:emd}

We have implemented an IMF filtering algorithm, known as Empirical Mode Decomposition (EMD), following Huang et al\cite{Huang1998,Huang2003}.  The IMFs and the residual trend line thus obtained are verified to be complete by simply summing them to recreate the signal.  The maximum relative error we have found is of the order $10^{-9}$ \%. 
\begin{figure}[htbp]
\begin{center}
\begin{tabular}{c}
\includegraphics[height=0.6\textwidth]{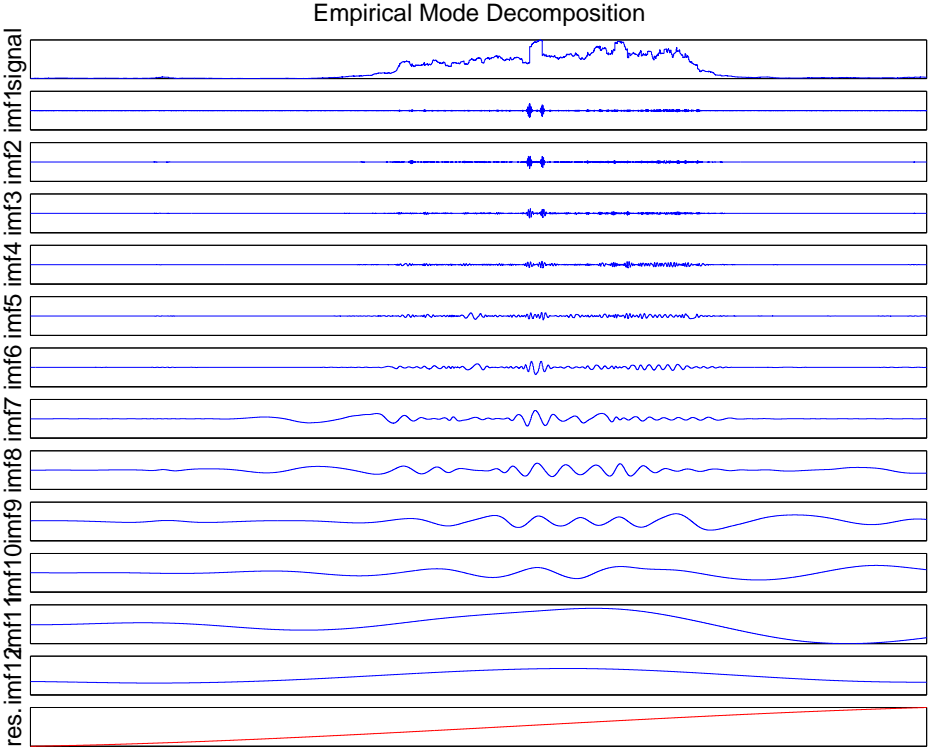}
\end{tabular}
\end{center}
\caption{\label{fig:imfexample} The IMFs found from a typical input signal (taken on 9-March-2006), with the signal itself shown at the top.  For convenience, we refer to the lowest order IMF as one with the fastest oscillation.  The bottom--most graph is the residual after removing all the IMFs and represents the overall trend.}
\end{figure}

\begin{figure}[htbp]
\begin{center}
\begin{tabular}{c}
\includegraphics[height=0.5\textwidth]{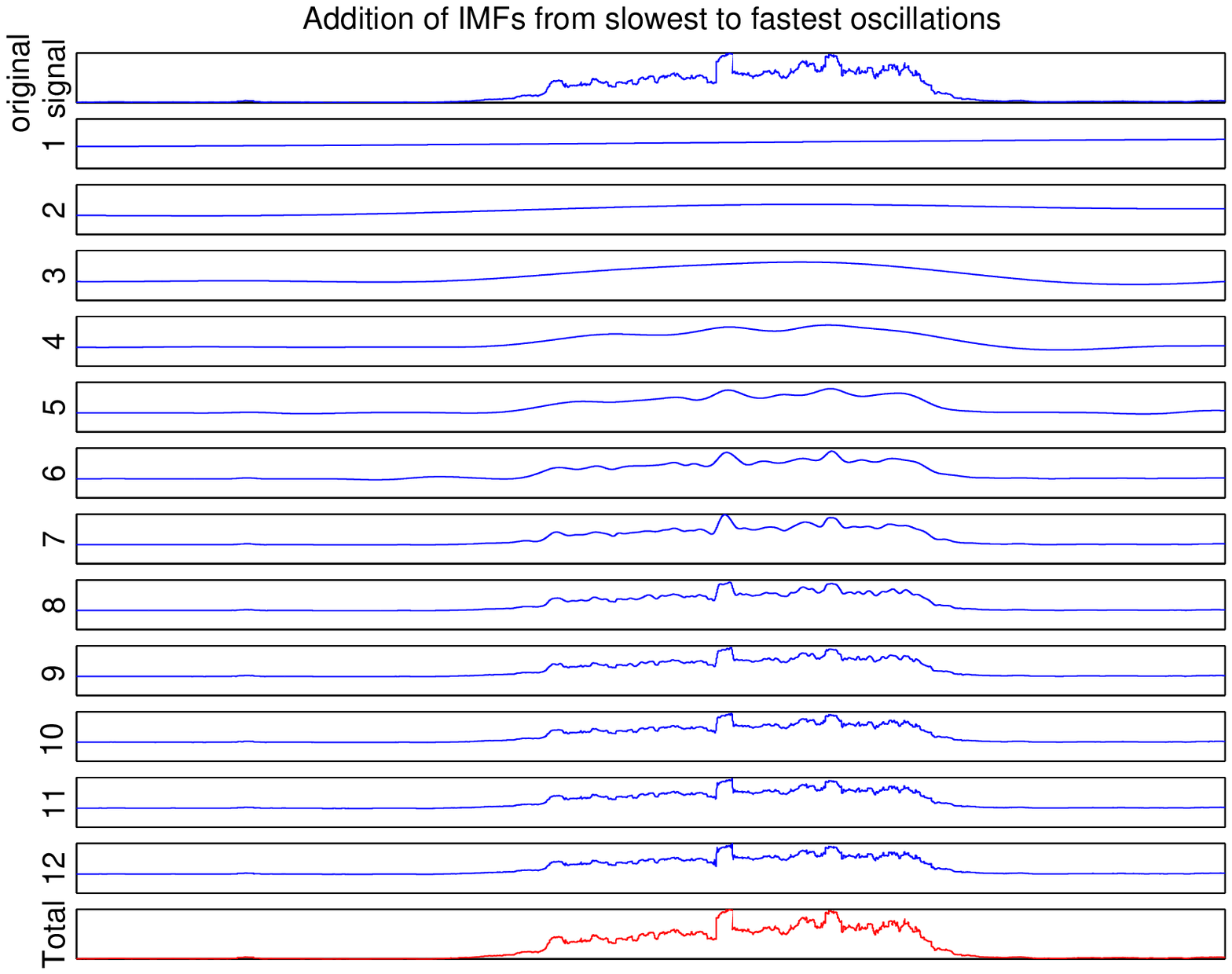} \\
\includegraphics[height=0.5\textwidth]{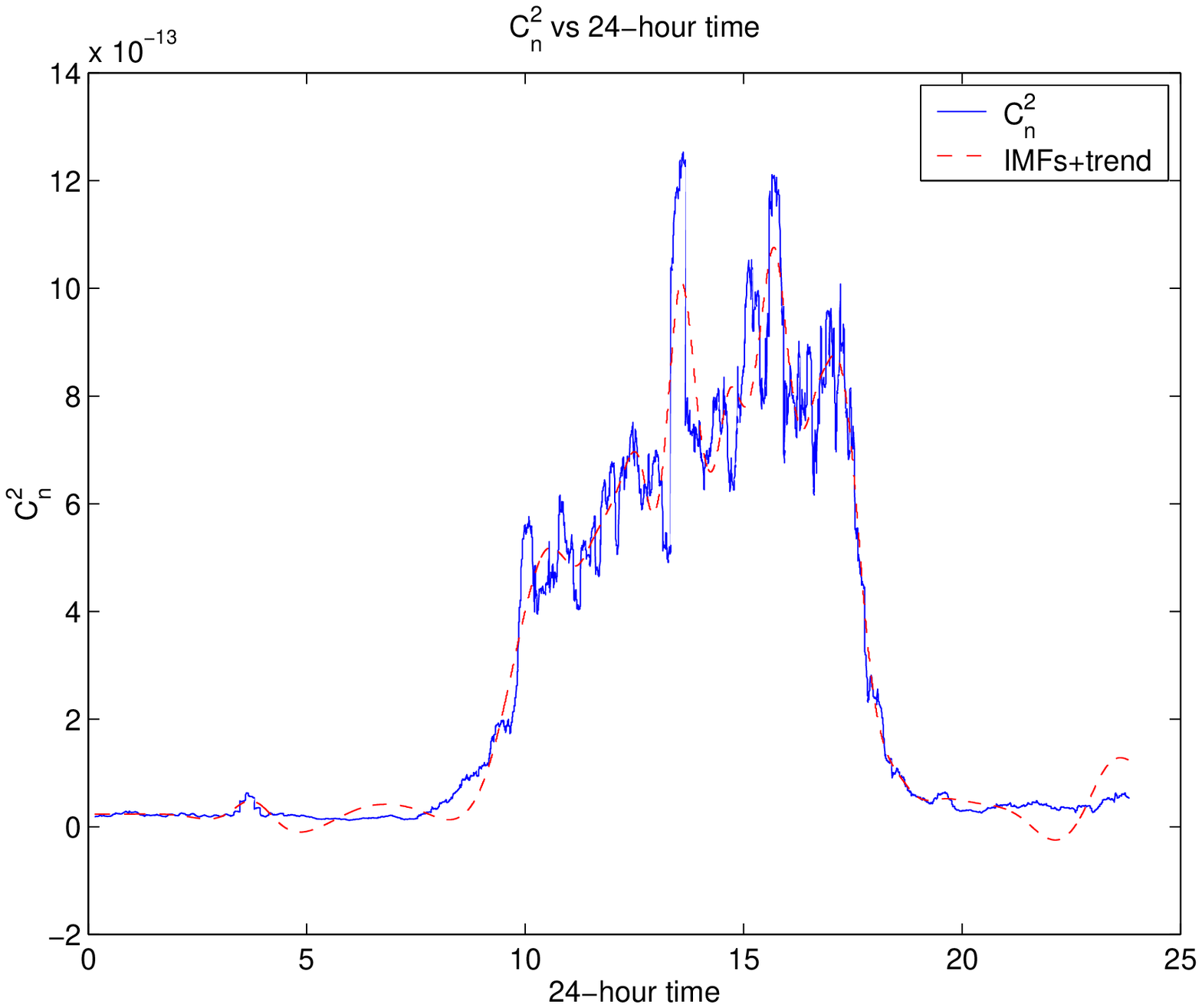}
\end{tabular}
\caption{\label{fig:solarfit} (a) Stepwise summation of the 9-March-2006 IMFs and trend line to recreate the original input signal.  (b) The sum of the trendline and the slowest 3 IMFs superimposed on the input signal.  The overshoot into negative values of $C_n^2$ is unphysical, and serves to demonstrate that the information content of the subset is incomplete.  Nevertheless, the fit does suggest that the IMFs represent an underlying physical process (primarily solar insolation).}
\end{center}
\end{figure}

\begin{figure}[htbp]
\begin{center}
\begin{tabular}{cc}
\includegraphics[height=0.35\textwidth]{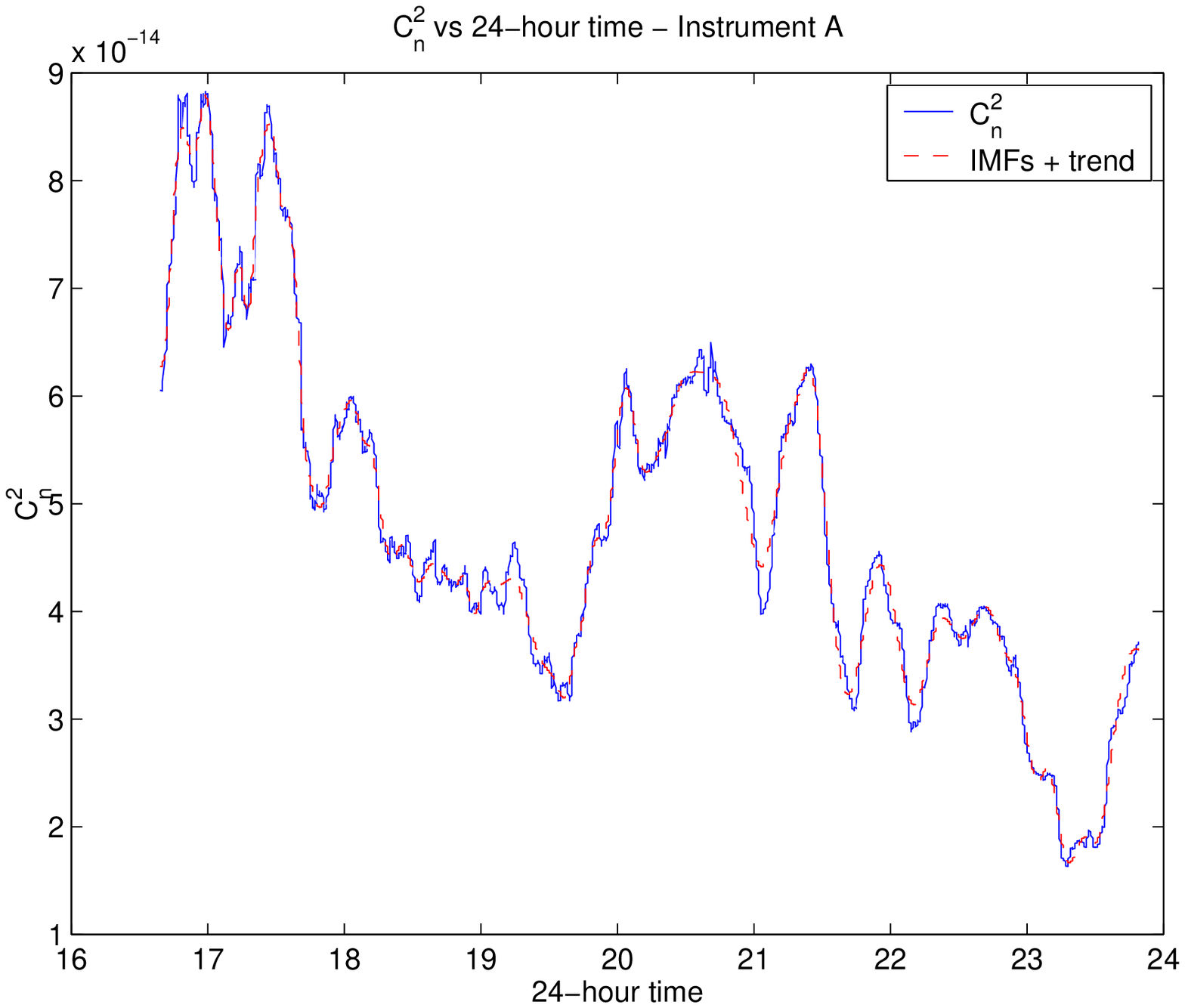} &
\includegraphics[height=0.35\textwidth]{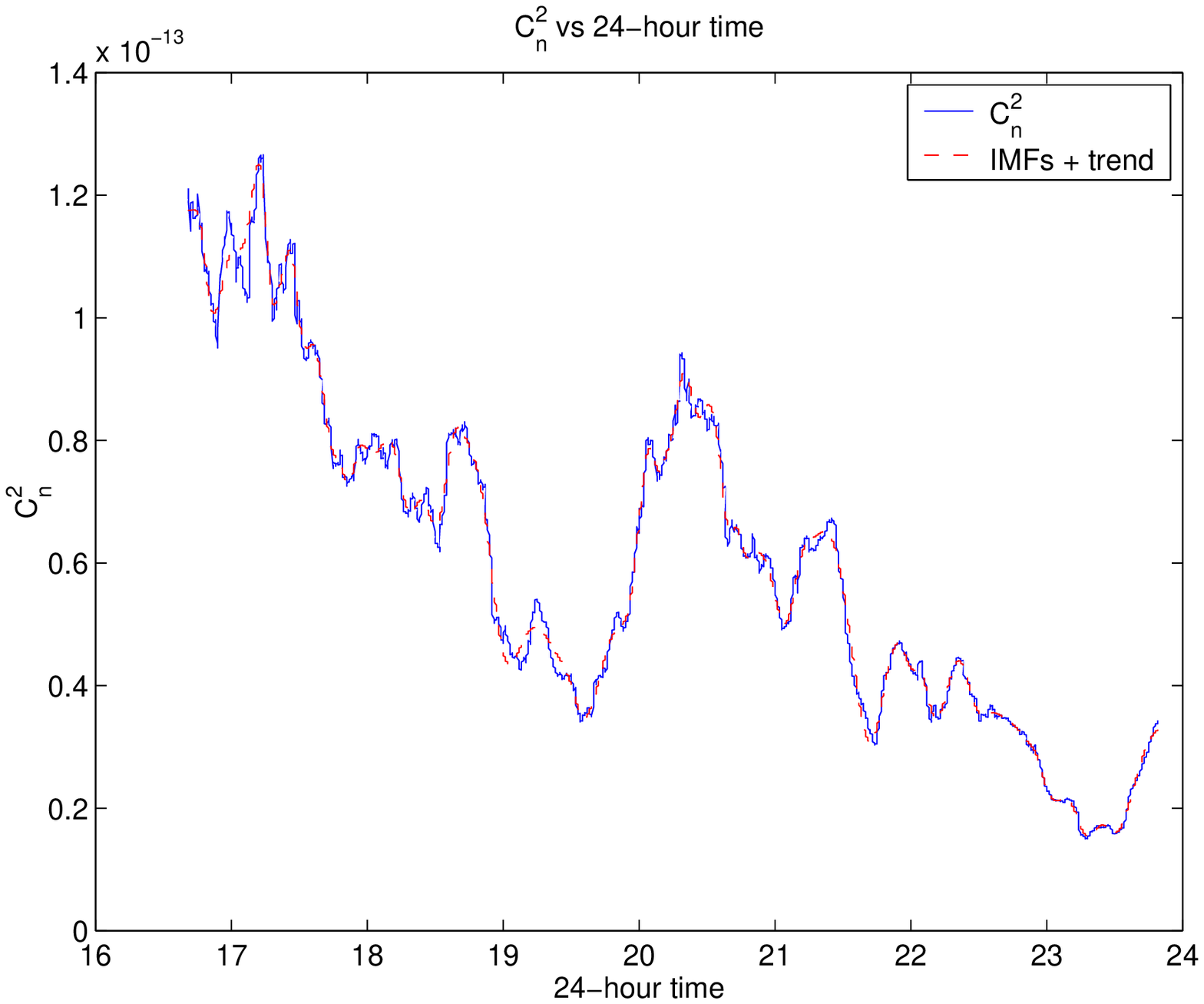}
\end{tabular}
\caption{\label{fig:evening} The 20-Feb-2006 $C_n^2$ and fit composed of the trend line and 3 highest order IMFs for both instruments (A on the left and B on the right).  Sunset was at 18:31 local time.}
\end{center}
\end{figure}

The IMFs show that in a very real sense the EMD method is acting as a filter bank, separating the more rapid oscillations from the slower oscillations.  It seems that a subset of the individual IMFs may be added to determine the effect of physical variables, as suggested in Figure \ref{fig:solarfit}.
In the absence of the major effects of solar insolation, the HHT technique reveals that the majority contribution to the $C_n^2$ signal lies in the highest order (slowest oscillation) IMFs, as can be seen from the extremely faithful fit to the data composed of the trend line and the 3 highest order IMFs shown in Figure \ref{fig:evening}.
As a guide to the significance of the various modes, we examined the energy in the IMFs and compared the energy in each mode to the energy distribution of red noise.  As an initial na\"{i}ve estimator, we take IMF1 to be representative of the noise contained in the data.  This is unlikely to be completely correct; probably a better noise estimator would be IMF1 - $<$IMF1$>$, where the angle brackets signify the mean value over the same temporal epoch (e.g. month or season).  We do not do this simply because we do not have sufficient data.

We define the red noise (random) time series to be an AR1 process
\begin{equation}
r(t_n) = \sigma E(t_n) + \rho r(t_{n-1})
\end{equation}
The terms are: 
\begin{eqnarray}
\sigma & := & \textrm{standard deviation of IMF1} \\ \nonumber
E & := & \textrm{uniform distribution of random numbers between 1 and -1} \\ \nonumber
t_n & := & \textrm{the } n\textrm{th timestep} \\ \nonumber
r & := & \textrm{the random time series} \nonumber
\end{eqnarray}
The IMFs of an ensemble of AR1 time series are generated and then this Monte Carlo is used to simulate the power distribution of the noise.  The power of each $C_n^2$ derived IMF is then compared to the noise power distribution to determine the mode's significance.
\begin{figure}[htbp]
\begin{center}
\begin{tabular}{c}
\includegraphics[height=0.5\textwidth]{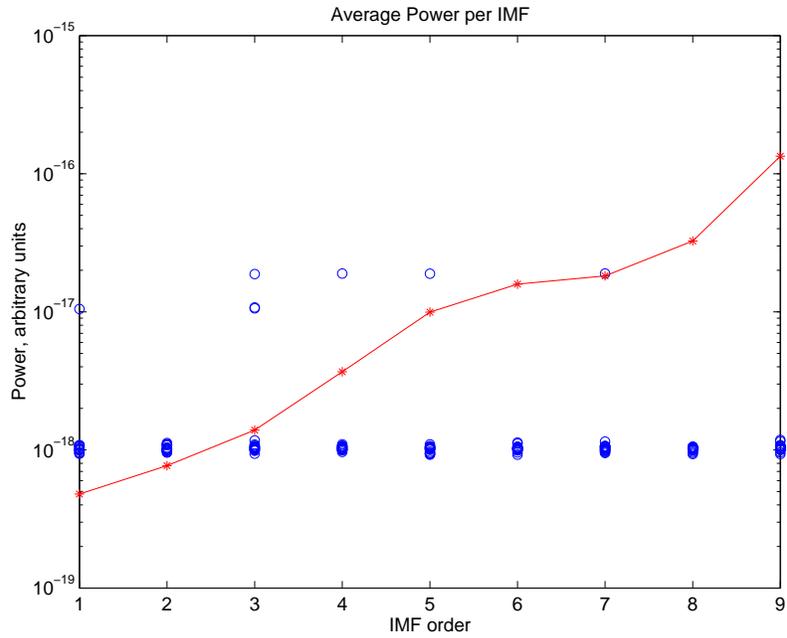}
\end{tabular}
\caption{\label{fig:montecarlo} Monte Carlo simulations show the significance of the $C_n^2$ (solid line) derived IMFs for a single instrument on the afternoon/evening of 20-Feb-2006, compared to the equivalent red noise power (open circles).}
\end{center}
\end{figure}
Figure \ref{fig:montecarlo} shows that of the IMFs, the first three lie at or below the median red noise power.  IMF4 to IMF9 lie above the median noise power, with the higher order IMFs being most significant.  We interpret this simulation to mean that IMF6-IMF9 are highly physically significant.

\section{HILBERT TRANSFORM OF IMFS}
\label{sect:hht}

Following the decomposition into IMFs of the original signal, the derived components can be Hilbert transformed to produce a time--frequency map or spectrum.
\begin{figure}[htbp]
\begin{center}
\begin{tabular}{c}
\includegraphics[height=0.5\textwidth]{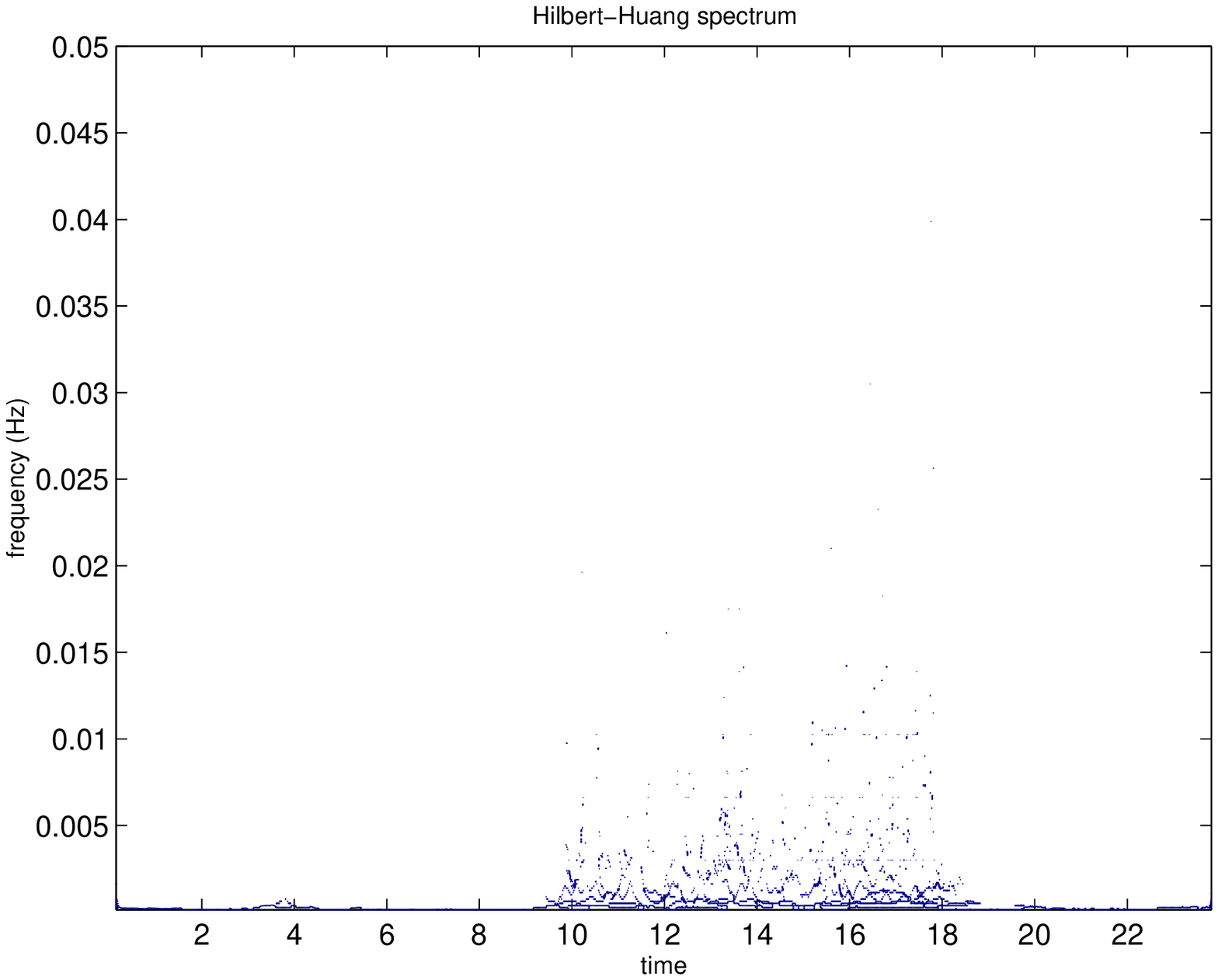}
\end{tabular}
\caption{\label{fig:hhspectrum} The Hilbert spectrum of the 9-March-2006 IMFs shown in Figure \ref{fig:imfexample} plotted as a series of contour lines.}
\end{center}
\end{figure}
Figure \ref{fig:hhspectrum} shows that during the hours when there is no solar insolation the $C_n^2$ energy is distributed in the lowest frequencies.  The highest frequencies sampled are only reached after the Sun is contributing energy into the lower atmosphere.  The discontinuous, filamentary aspect of the plot indicates a large number of phase dropouts which shows that the data are non--stationary.

\begin{figure}[htbp]
\begin{center}
\begin{tabular}{c}
\includegraphics[height=0.5\textwidth]{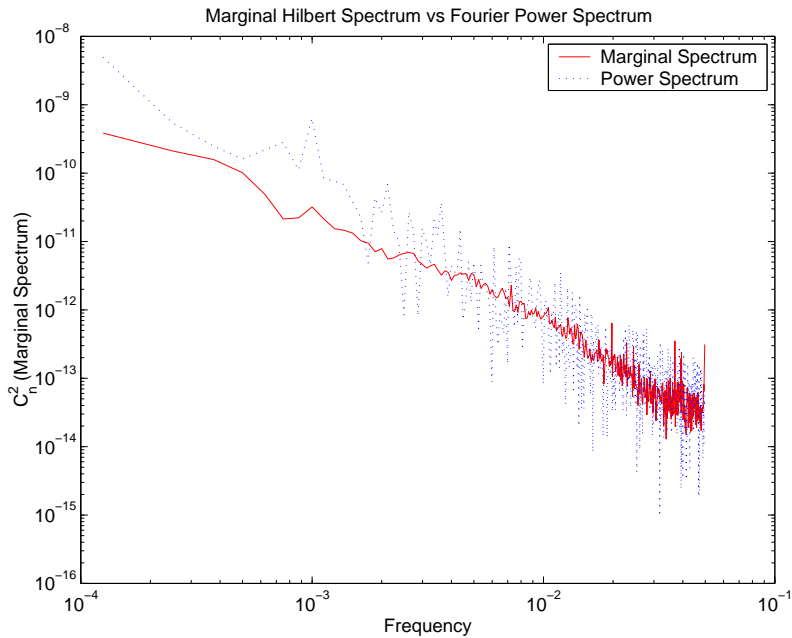}
\end{tabular}
\caption{\label{fig:margfft} The Hilbert Marginal spectrum (solid line) of the 9-March-2006 IMFs compared to the Fourier Power Spectrum (dotted line) of the same data.  The Power Spectrum has been shifted so that its maximum frequency coincides with that of the Marginal Spectrum.  Note also that the units of the ordinate axis are arbitrary for the Fourier Power Spectrum.}
\end{center}
\end{figure}

We are also able to find a Marginal Spectrum by integrating the Hilbert spectrum across time.  The Marginal Spectrum shown in Figure \ref{fig:margfft} clearly suffers from less leakage into the high frequencies than the Power Spectrum.  The interpretation of both spectra are quite different: the Fourier Power Spectrum indicates that certain frequencies exist throughout the entire signal with a given squared amplitude.  The Marginal Spectrum, on the other hand, describes the probability that a frequency exists at some local time point in the signal.
It is clear that the two spectra have a difference in the standard deviations: the logarithm of the Marginal Spectrum has a standard deviation of 1.88 compared to the logarithm of the Fourier Power Spectrum, whose standard deviation is 2.24.

Applying a Kolmogorov--Smirnov test to the two spectra returns a P--value of 0 and a cumulative distribution function distance of 1, indicating that the data sets represent different distributions, as we might guess from their different gradients.  We may therefore state that the spectra are unrelated and we argue on the basis of non--stationarity that the Fourier Power Spectrum has little, if any, physical meaning.
\section{INSTRUMENTS AND DATA REDUCTION} 
\label{sect:instruments}

The $C_n^2$ data used in this study was collected during 2006 at the University of Puerto Rico, Mayag\"{u}ez Campus, on the rooftop of the Physics Department.

The data were obtained with two commercially available scintillometers (model LOA-004) from Optical Scientific Inc, co--located such that the transmitter of system 1 was next to the receiver of system 2.

Each LOA-004 instrument comprises of a single modulated infrared transmitter whose output is detected by two single pixel detectors.  For these data, the separation between transmitter and receiver was just under 100-m.  The sample rate was set to 10 seconds, so that each $C_n^2$ point was found from a 10 second time average.
The path integrated $C_n^2$ measurements are determined by the LOA instruments by computation from the log--amplitude scintillation ($C_\chi(r)$) of the two receiving signals\cite{Ochs1979,Wang2002}.  The algorithm for relating $C_\chi(r)$ to $C_n^2$ is based on an equation for the log--amplitude covariance function in Kolmogorov turbulence by Clifford {\em{et al.}}\cite{Clifford1974}

\begin{figure}[htbp]
\begin{center}
\begin{tabular}{cc}
\includegraphics[height=0.35\textwidth]{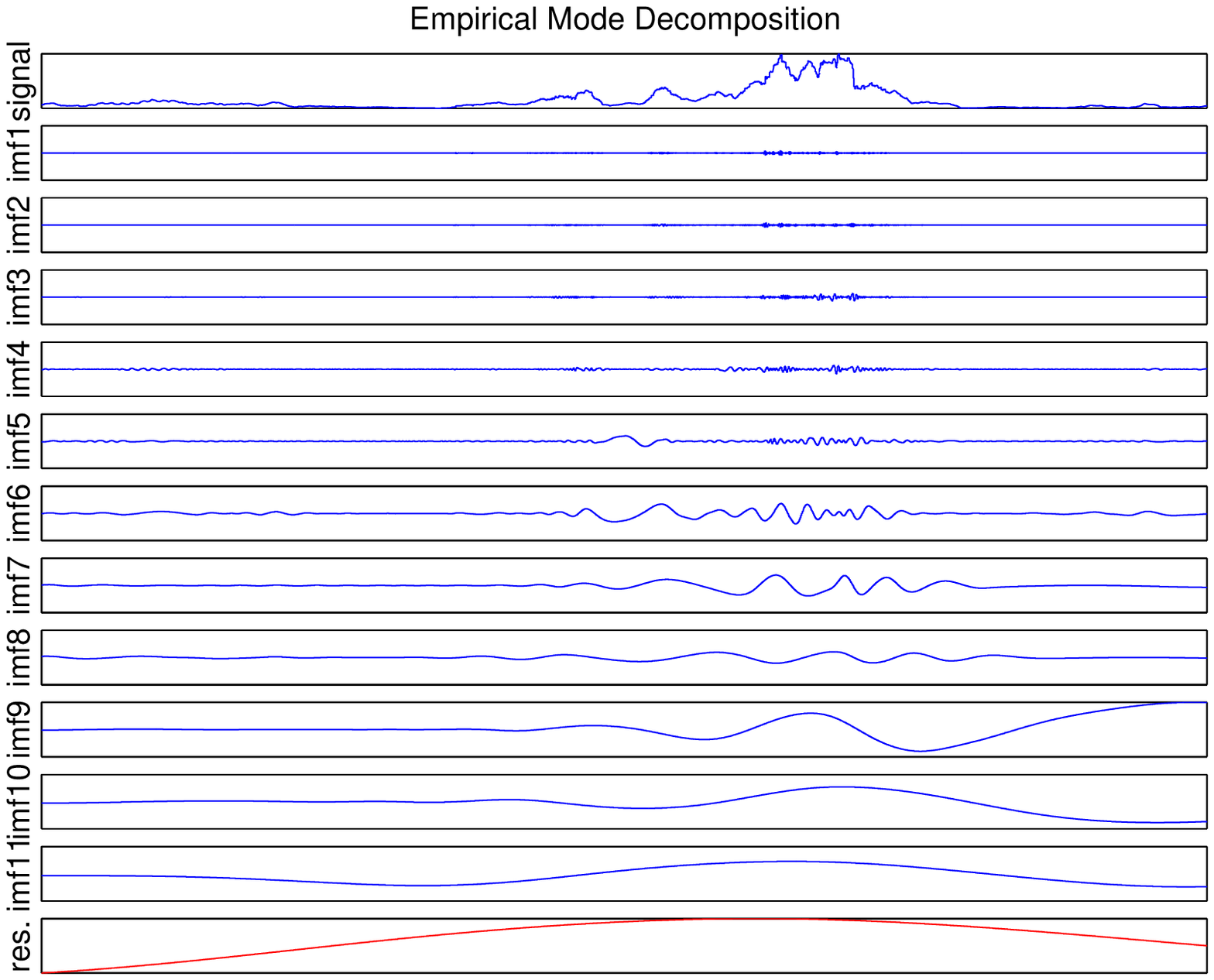} &
\includegraphics[height=0.35\textwidth]{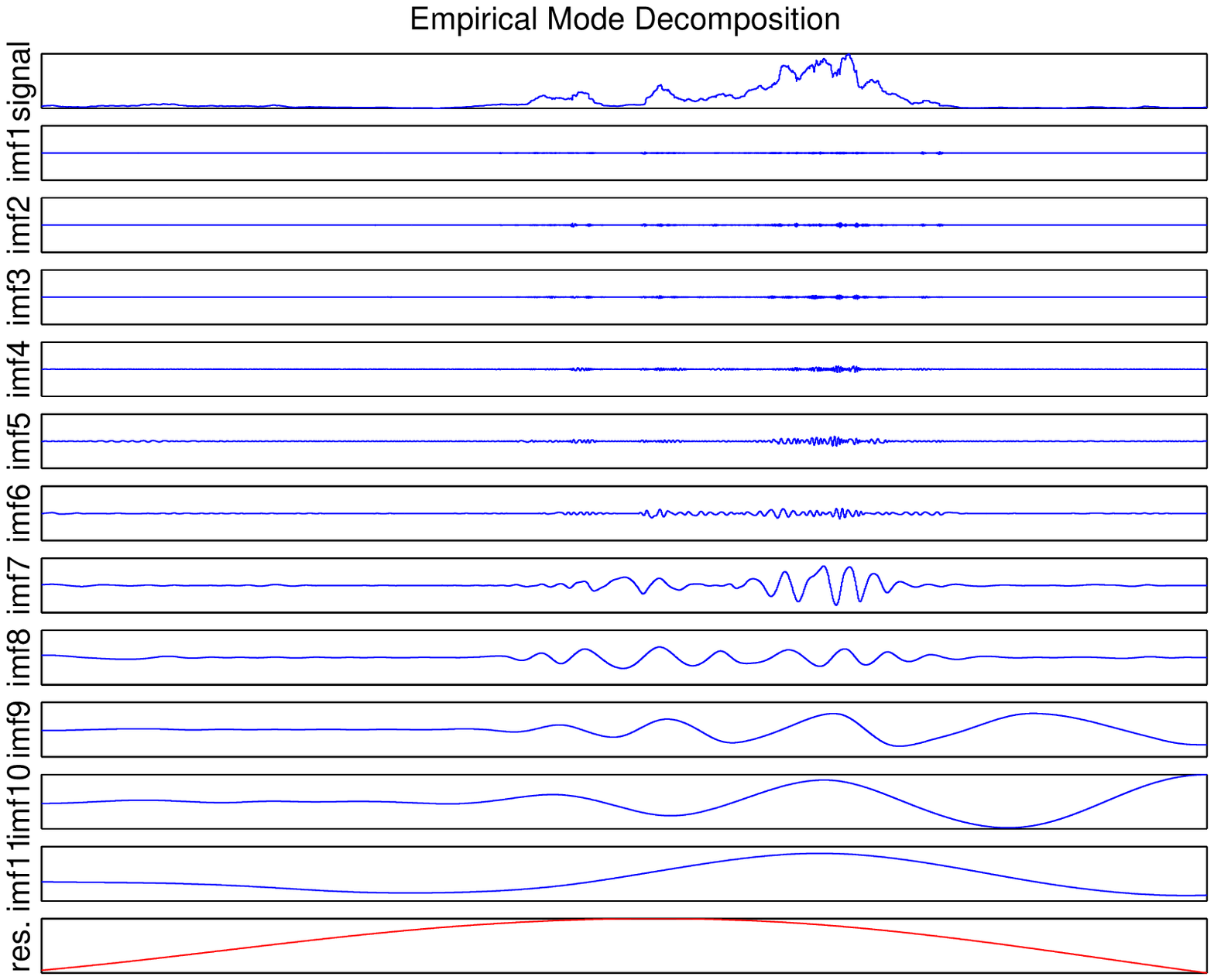}
\end{tabular}
\caption{\label{fig:comparisonimfs} The IMFs extracted from each instrument for the same day (3-March-2006), beginning and ending at midnight (A on the left and B on the right).}
\end{center}
\end{figure}

\begin{figure}[htbp]
\begin{center}
\begin{tabular}{c}
\includegraphics[height=0.5\textwidth]{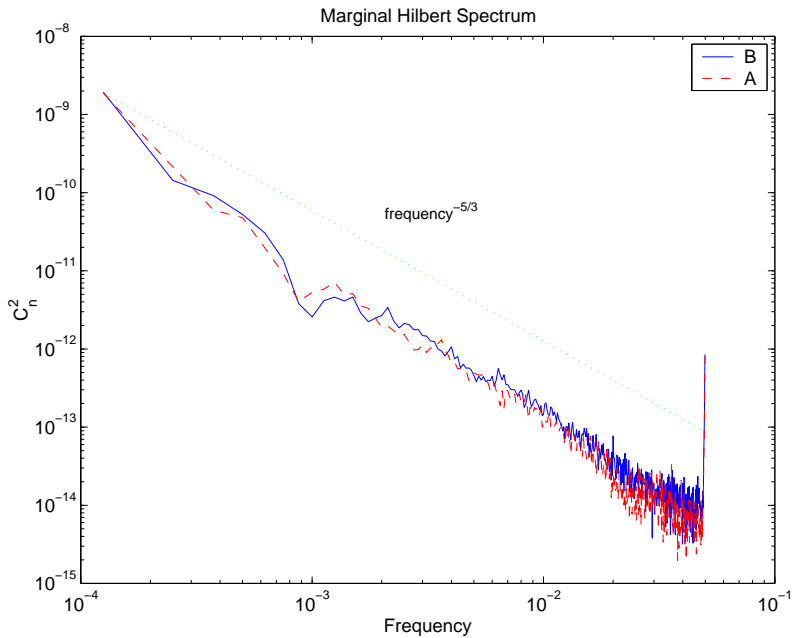}
\end{tabular}
\caption{\label{fig:comparisonmarginals} The 3-March-2006 Marginal Spectra of both instruments with a comparison frequency$^{-5/3}$ (Kolmogorov Spectrum) trend line.}
\end{center}
\end{figure}

The data was collected by dedicated PCs, one per instrument.  During analysis, the data were smoothed by a 120 point (10 minute) boxcar rolling average.  This value was chosen for future ease of comparison with local weather station data, sampled at one reading per 10 minutes.  Figure \ref{fig:comparisonimfs} compares the extracted IMFs from a single day, from midnight to midnight.  There are no data dropouts in the time signal for instrument A, while instrument B is 99.81\% \ valid.  A visual examination reveals that the measured $C_n^2$ functions are very similar and both instruments have 11 IMFs.  Differences are to be found in the IMFs themselves.  
In Figure \ref{fig:comparisonmarginals} we show the Hilbert Marginal Spectra derived from the IMFs together with a Kolmogorov Power Spectrum trend scaled to start coincident with the Marginal Spectra.  Both Marginal Spectra follow each other fairly well, with similar frequency probabilities.
Applying a Kolmogorov--Smirnov test to the two Marginal Spectra data sets gives the same means and standard deviations.  This is indicative of a P--value of 1 and a cumulative distribution function distance of 0, so we may conclude that the instrument outputs have come from exactly the same distribution and are statistically identical.  A further confirmation can be found by calculating the Hilbert phase difference between the two Marginal Spectra.  Such a test displays phase synchronization, or lack thereof.  In this case we find a phase difference of zero, so that the spectra are perfectly in phase.

\newpage
\section{CONCLUSIONS}

We have presented the results of applying the Hilbert Huang Transform to $C_n^2$ time series data.  When used to compare the outputs of two of the same model of commercial scintillometer, we have been able to demonstrate that they provide identical output in terms of their Hilbert Marginal Spectra.  It is clear that the HHT technique is a very useful tool in the analysis of non--stationary turbulence data and promises much in terms of understanding the nature of optical turbulence.

\acknowledgments     
MPJLC would like to thank Norden Huang for introducing him to the Hilbert Huang Transform.  Thanks also are due to Sergio Restaino and Christopher Wilcox for making available the scintillometers and providing data processing software.


\bibliography{bib}   
\bibliographystyle{spiebib}   

\end{document}